# Opinion Mining In Hindi Language: A Survey

Richa Sharma[1], Shweta Nigam[2] and Rekha Jain[3]

[1,2]M.Tech Scholar, Banasthali Vidyapith, Rajasthan, India
[3]Assistant Professor, Banasthali Vidyapith, Rajasthan, India

*ABSTRACT*

*Opinions are very important in the life of human beings. These Opinions helped the humans to carry out the decisions. As the impact of the Web is increasing day by day, Web documents can be seen as a new source of opinion for human beings. Web contains a huge amount of information generated by the users through blogs, forum entries, and social networking websites and so on To analyze this large amount of information it is required to develop a method that automatically classifies the information available on the Web. This domain is called Sentiment Analysis and Opinion Mining. Opinion Mining or Sentiment Analysis is a natural language processing task that mine information from various text forms such as reviews, news, and blogs and classify them on the basis of their polarity as positive, negative or neutral. But, from the last few years, enormous increase has been seen in Hindi language on the Web. Research in opinion mining mostly carried out in English language but it is very important to perform the opinion mining in Hindi language also as large amount of information in Hindi is also available on the Web. This paper gives an overview of the work that has been done Hindi language.*

## 1. INTRODUCTION

Most of the peoples now a days would like to share their feelings ,experiences and opinions on the Web.Now people commonly use blogs, forums, e-news, reviews channels and the social networking platforms such as Facebook, Twitter, to express their views and opinions. The World Wide Web plays a crucial role in gathering public opinion. These opinions are very helpful for the business organizations, as they would know about the sentiments/opinions of their user about their products and also helps the customers in taking the decision. Large amount of user content data is generated on the Web every day, thus mining the data and identifying user sentiments, wishes, likes and dislikes is one of an important task.

Sentiment Analysis is a natural language processing task that deals with finding orientation of opinion in a piece of text with respect to a topic [8]. It mines the information from various text forms such as reviews, news, and blogs and classifies them on the basis of their polarity as positive, negative or neutral. It focuses on categorizing the text at the level of subjective and objective nature. Subjectivity indicates that the text contains/bears opinion content whereas Objectivity indicates that the text is without opinion content [9].

Some examples-

**Subjective-** शाहरुख और काजोल की यह फिल्म अच्छी है। (this sentence has an opinion, it talks about the movie and the writer's feelings about "अच्छी" and hence it's subjective).The subjective text can be further categorized into 3 broad categories based on the sentiments expressed in the text.

1. **Positive**- यह फिल्म अच्छी है।
2. **Negative**- यह होटल बहुत खराब है।
3. **Neutral**- मुझे दोपहर तक भूख लगने लगती है। (this sentence has user's views, feelings hence it is subjective but as it does not have any positive or negative polarity so it is neutral.)

**Objective**- इस फिल्म में शाहरुख और काजोल है। (This sentence is a fact, general information rather than an opinion or a view of some individual and hence its objective).

### 1.1 Components of Opinion Mining

There are mainly three components of Opinion Mining, these are [9]:

- **Opinion Holder:** Opinion holder is the holder of a particular opinion; it may be a person or an organization that holds the opinion. In the case of blogs and reviews, opinion holders are those persons who write these reviews or blogs.
- **Opinion Object:** Opinion object is an object on which the opinion holder is expressing the opinion.
- **Opinion Orientation:** Opinion orientation of an opinion on an object determines whether the opinion of an opinion holder about an object is positive, negative or neutral.

For example "इस मोबाइल का कैमरा अच्छा है।". In this review, the person who has written this review is the Opinion Holder. Opinion object here is the कैमरा of the mobile and the opinion word is "अच्छा" which is positively orientated. Semantic orientation is a task of determining whether a sentence has either positive, negative orientation or neutral orientation [6] [23].

### 1.2 Levels of Opinion Mining

Research in the field of sentiment analysis is done at various levels which are as follows [9]-

- **Document Level**- The document level classifies the whole document as a single polarity positive, negative or neutral.
- **Sentence Level** -The sentence level analyze the documents at sentence level. The sentences are analyzed individually and classified as positive, negative or neutral.
- **Aspect Level**- The aspect level analyze going much deeper and deals with identifying the features in a sentence for a given document and analyze the features and classify them accordingly as positive, negative or neutral.

Most of the work in Opinion mining has been done in English language; Very little attention has been paid in the direction of sentiment analysis for other languages. As the internet is reaching to more and more people within the world, there is tremendous increase in Web content of other languages because people feel comfortable with their native language. For the last few years there has been an enormous increase in the Hindi content on the web. Hindi is the 4th largest spoken language and has 490 million speakers across the world majority of whom are from India (Wikipedia). With Unicode (UTF-8) standards for Indian languages introduced, web pages in Hindi language have increased at a very fast rate. There are many websites which provide information in Hindi, ranging from various news websites such as[15][16] for sites providing information regarding the culture, music, entertainment and other aspects of arts such as [17][18][19] There are various weblogs which are written in Hindi. Large amount of Hindi content is available on the web, so it is necessary to mine this large amount of information and extracting the opinions from this data which helps the users and the organizations in taking decisions. This paper discusses about opinion mining in Hindi language. The paper is organized into the following sections: Section 2 discusses existing research work that has been performed in this language. Section 3 explains various challenges in sentiment analysis in Hindi language. The last section concludes the study.

## 2. EXISTING RESEARCH WORK

In the field of opinion mining a small amount of work has been done in Hindi language. The very first research has been done in Hindi, Bengali and Marathi language. Amitava Das and Bandopadhya [1], developed sentiwordnet for Bengali language. Word level lexical-transfer technique have been applied to each entry in English SentiWordNet using an English-Bengali Dictionary to obtain a Bengali SentiWordNet. 35,805 Bengali entries has been returned by their experiment.

Das and Bandopadhya [2], devised four strategies to predict the sentiment of a word. In the first approach, an interactive game is proposed by them which turn annotated the words with their polarity. In Second approach, they use Bi-Lingual dictionary for English and Indian Languages to determine the polarity. In Third approach, they use wordnet and use synonym and antonym relations, to determine the polarity .In Fourth approach; learning from pre-annotated corpora takes place to determine the polarity.

Dipankar Das and Bandopadhya[11], emotional expressions are identified by them in Bengali corpus. Emotional expressions are identified on the basis of emotional components such as holders, intensity and topics. For sentence level annotation, words were classified by them in six emotion classes and with three types of intensities.

Joshi et al. [3] proposed a fallback strategy for Hindi language. This strategy follows three approaches: In-language Sentiment Analysis, Machine Translation and Resource Based Sentiment Analysis. They developed a lexical resource, Hindi SentiWordNet (HSWN) based on its English format. By using two lexical resources (English SentiWordNet and English-Hindi WordNet Linking [20]) H-SWN (Hindi-SentiWordNet) was created by them. By using Wordnet linking , words in English SentiWordNet were replaced by equivalent Hindi words to get H-SWN. The final accuracy achieved by them is 78.14.

By using a graph based method Bakliwal et al.[4]created lexicon .They determine that by using simple graph traversal how the synonym and antonym relations can be used to generate the subjectivity lexicon. Their proposed algorithm achieved approximately 79% accuracy on classification of reviews and 70.4% agreement with human annotated.

Mukherjee et al. [26] showed that the incorporation of discourse markers in a bag-of-words model improves the sentiment classification accuracy by 2 - 4%. Bakliwal et al. [5] proposed a method to classify Hindi reviews as positive or negative. They devised a new scoring function and test on two different approaches. They also used a combination of simple N-gram and POS Tagged N-gram approaches.

Ambati et al. [7] proposed a novel approach to detect errors in the treebanks. This approach can significantly reduce the validation time. They tested it on Hindi dependency treebank data and were able to detect 76.63% of errors at dependency level.

Piyush Arora et al. [25] proposed a graph based method to build a subjective lexicon for Hindi language,using WordNet as a resource. They build a subjective lexicon for Hindi language with dependency on WordNet. They initially build small seed list of opinion words and by using WordNet, synonyms and antonyms of the opinion words were determined and added to the seedlist .They traverse Wordnet like a graph where every word in a Wordnet considered as a node,which is connected to their synonyms and antonyms. They achieved 74% accuracy on classification of reviews and 69% accuracy is achieved in agreement with human annotators for Hindi.

Harshada Gune et al. [13] perform shallow parsing on Marathi language. They build a Marathi shallow parser which consists of Marathi POS tagger and Chunker. In their proposed system, morphological analyzer provides ambiguity and suffix information for generating a rich set of features. Generated features are then applied to the CRF based engine which couples them with other elementary features for training a sequence labeller. Verb Group Identifier (VGI) used by the POS tagger for correcting the output of the CRF based sequence labeller. 50% accuracy is achieved by their system.

Namita mittal et al.[22]developed an efficient approach based on negation and discourse relation to identifying the sentiments from Hindi content .They developed an annotated corpus for Hindi language and improve the existing Hindi SentiWordNet (HSWN) by incorporating more opinion words into it. Then they devised the rules for handling negation and discourse that affect the sentiments expressed in the review. Their proposed algorithm achieved approximately 80% accuracy on classification of reviews.

## 3. APPROACHES

Two types of techniques used in opinion mining:
- Supervised learning
- Unsupervised learning

### 3.1 Supervised learning

Supervised learning is a machine learning approach[6] , two types of document sets are required in this approach: training set and test set. A training set trained the classifier with the help of training data and a test set test the classifier by giving the test data. Large numbers of machine learning techniques are available, for example, Naïve Bayes classification, and Support Vector Machines which classifies the opinions

**Naïve Bayes:** A Naive Bayes Classifier is based on Bayes' theorem and is particularly used when the input dimensions are high. Naïve Bayes classification is a text classification approach that assigns the class c to a given document *d* given in eq (1).

$$c* = \arg max_c P(c|d) \qquad \text{eq (1)}$$

Where P(c|d) is the probability of instance d being in class c[24].

International Journal in Foundations of Computer Science & Technology (IJFCST), Vol.4, No.2, March 2014

**Support Vector Machine:** Support Vector Machine Classifier constructs N-dimensional hyperplane represented by vector $\vec{\omega}$ which separates data into two categories. SVM takes the input data and for each input it predicts the class. SVM can be seen as a constrained optimization problem, in which class $c_j$ {1, −1} corresponds to either positive or negative class that belongs to document $d_j$, the solution can be written as in eq (2)

$$\vec{\omega} = \sum_j \alpha_j\ c_{j_{\vec{d_j}}},\ \alpha_j \geq 0 \qquad \text{eq (2)}$$

Where $\vec{\omega}$ is a vector, $c_j$ is a class and $d_j$ is a document [14].

### 3.2 Unsupervised learning

Unsupervised learning, models a set of inputs, like clustering, labels are not known during training. [10]. In Unsupervised learning classification is done by comparing the opinion of a given text against word lexicons whose sentiment values are determined prior to their use.

With the help of word lexicons the sentiment orientation of the documents are determined. Sentiment orientation is an unsupervised learning because no prior training is required for opinion mining [23]. For example, to find whether the document possesses positive opinion or negative opinion, all the opinion words in the document are first extracted and their polarity is determined with the help of word lexicon which contains the opinion words with polarity. When the polarity of all the opinion words is determined then it is checked whether the document contain more positive word or more negative words. If the document contains more positive words then negative words, the sentiment orientation of the document is positive otherwise it is negative. The part-of-speech (POS) tags are used to determine the opinion words. A Part-Of-Speech Tagger (POS Tagger) is a software that reads the input text and assigns parts of speech to each word in input text [10]. For example, Suppose the sentence is, **"This book is good"**, POS tagger gives the output **"This/DT book/NN is/VBZ good/JJ /. /"**

There are two types of technique in unsupervised learning:

**Corpus-based techniques**: In this method a seed set of opinion words whose polarity are known is used and co-occurrence patterns are used to determine the new opinion words and their polarity in a large corpus. [27].

**Dictionary-based techniques:** In this method, initially a seed list of opinion words with their polarity is constructed manually and then with the help of available lexicographical resources like WordNet, Senti Wordnet etc, synonyms and antonyms of respective words are determined to expand the seed list. [21]. WordNet is a large lexical database in which words are grouped into sets of synonyms, antonyms and hyponyms each define a different concept [12].

## 4. CHALLENGES

Challenges while dealing/working with Hindi language are as follows [7]-

1) **Word Order-** Word arrangement in a sentence plays an important role in identifying the subjective nature of the text. Hindi is a free order language i.e. the subject, object and verb can come in any order whereas English is a fixed order language i.e. subject followed by a verb and followed by an object. Word order plays a vital role in deciding the polarity of a text, in the text same set of words with slight variations and changes in the word order affect the polarity aspect.

International Journal in Foundations of Computer Science & Technology (IJFCST), Vol.4, No.2, March 2014

2) **Morphological Variations-** Handling the morphological variations is also a big challenge for Hindi language. Hindi language is morphologically rich which means that lots of information is fused in the words as compared to the English language where we add another word for the extra information.
3) **Handling Spelling Variations-** In the Hindi language, the same word with same meaning can occur with different spellings, so it's quite complex to have all the occurrences of such words in a lexicon and even while training a model it's quite complex to handle all the spelling variants.
4) **Lack of resources-** the lack of sufficient resources, tools and annotated corpora also adds to the challenges while addressing the problem of sentiment analysis especially when we are dealing with Non-English languages.
5) **Co reference resolution-** It is a problem of determining multiple expressions that refer to the same thing. For example "राम ने खाना खाया। वह सोने चला गया।" "वह" in the second sentence refers to राम that is an entity. It is important to recognize these co reference relationships for aspect-based sentiment analysis.

## 5. CONCLUSION

Opinion Mining is an emerging research field and is very important because human beings are largely dependent on the web nowadays. The rise in user-generated content in Hindi language across various genres- news, culture, arts, sports etc. has open the data to be explored and mined effectively, to provide better services and facilities to the consumers. Opinion Mining has large application areas like Shopping, where websites like flipkart.com, amazon.com etc. Allow customers to express their opinions on their websites which helps other customers to decide whether to buy the product or not. Entertainment, where the people can easily see the critics and viewer's reviews of their favourite movies and shows online. Marketing, now every organization and private firms allow their customers to write the reviews related to their products on their websites which eliminate the needs to conduct surveys.

Large amount of work in opinion mining has been done in English language, as English is a global language, but there is a need to perform opinion mining in other languages also. Large amount of other languages contents are available on the Web which needs to be mined to determine the opinion. Hindi is a national language of India, large amount of Hindi content is available on the Web. Google and Yahoo search engine also provide the services in Hindi language; one can see write and search the information in Hindi on these search engines so Opinion Mining in Hindi language is required. From the last few years researchers has performed opinion mining in Hindi language. In this paper an overview of the Hindi based Opinion Mining has given, based on existing researches that has been performed in Hindi language. Techniques and several challenges of Hindi based Opinion Mining are also discussed. But performing opinion mining in Hindi language is not an easy task, because the nature of Indian languages varies a great deal in terms of the script, representation level and linguistic characteristics, etc. To understand the behaviour of Indian languages, large amount of work needs to be done in the field of opinion mining for Hindi language.

## REFERENCES

[1] A. Das and S. Bandyopadhyay,(2010),"SentiWordNet for Bangla". Knowledge Sharing Event-4: Task,Volume 2
[2] A. Das and S. Bandyopadhyay, (2010),"SentiWordNet for Indian Languages",Asian Federation for Natural Language Processing(COLING), China ,Pages 56-63
[3] A. Joshi, B. A. R, and P. Bhattacharyya.(2010)," A fall-back strategy for sentiment analysis in Hindi: a case study" In International Conference On Natural Language Processing (ICON).46

International Journal in Foundations of Computer Science & Technology (IJFCST), Vol.4, No.2, March 2014[4] Akshat Bakliwal, Piyush Arora, Vasudeva Varma,(2012)"Hindi Subjective Lexicon : A Lexical Resource For Hindi Polarity Classification".In Proceedings of the Eight International Conference on Language Resources and Evaluation (LREC) .
[5] Akshat Bakliwal, Piyush Arora, Ankit Patil, Vasudeva Varma,(2011),"Towards Enhanced Opinion Classification using NLP Techniques" In Proceedings of the Workshop on Sentiment Analysis where AI meets Psychology (SAAIP), IJCNLP, pages 101–107, 2011
[6] B. Pang, L. Lee, and S. Vaithyanathan,(2002),"Thumbs up? Sentiment classification using machine learning techniques" In Proceedings of the 2002 Conference on Empirical Methods in Natural Language Processing (EMNLP), pages 79–86.
[7] Bharat R. Ambati, Samar Husain, Sambhav Jain, DiptiM.Sharma, Rajeev Sangal,(2010) "Two Methods to Incorporate Local Morph Syntactic Features in Hindi Dependency Parsing" In Proceedings of the NAACL HLT 1st Workshop on Statistical Parsing of Morphologically-Rich Languages, pages 22–30.
[8] Bo Pang, Lillian Lee,(2008)"Opinion mining and sentiment analysis". Foundations and Trends in Information Retrieval, Vol. 2(1-2):pp. 1–135.
[9] Bing Liu,(2012), "Sentiment Analysis and Opinion Mining, Morgan & Claypool Publishers".
[10] Christopher D. Manning ,"Part-of-Speech Tagging from 97% to 100%: Is It Time for Some Linguistics?",Published in CICLing'11 Proceedings of the 12th international conference on Computational linguistics and intelligent text processing - Volume Part I.
[11] D. Das and S. Bandyopadhyay,(2010)"Labeling emotion in bengali blog corpus a fine grained tagging at sentence level", In Proceedings of the Eighth Workshop on Asian Language Resouces, pages 47–55, Beijing, China, Coling Organizing Committee.
[12] George A. Miller, Richard Beckwith, Christiane Fellbaum, Derek Gross, and Katherine Miller (1990)," Introduction to WordNet: An On-line Lexical Database (Revised August 1993) International Journal of Lexicography" 3(4):235-244.
[13] Harshada Gune, Mugdha Bapat, Mitesh Khapra and Pushpak Bhattacharyya,(2010),"Verbs are where all the action lies: Experiences of shallow parsing of a morphologically rich language", In Proceedings of COLING , Beijing,China.
[14] H. Cui, V. Mittal, and M. Datar, (2006)"Comparative experiments on sentiment classification for online product reviews," presented at the proceedings of the 21st national conference on Artificial intelligence - Volume 2, Boston, Massachusetts.
[15] http://dir.hinkhoj.com/
[16] http://bbc.co.uk/hindi
[17] http://www.webdunia.com/
[18] http://www.virarjun.com/
[19] http://www.raftaar.in/
[20] Karthikeyan,"Hindi english wordnet linkage".
[21] M. Hu and B. Liu, "Mining and summarizing customer reviews," presented at the Proceedings of the tenth ACM.
[22] Namita Mittal,Basant Agarwal,Garvit Chouhan,Nitin Bania,Prateek Pareek,(2013), "Sentiment Analysis of Hindi Review based on Negation and Discourse Relation"in proceedings of International Joint Conference on Natural Language Processing, pages 45–50,Nagoya, Japan, 14-18 .
[23] P. Turney, "Thumbs up or thumbs down? Semantic orientation applied to unsupervised classification of reviews," Proceedings of the Association for Computational Linguistics.
[24] Pedro Domingos and Michael J. Pazzani,(1997),"On the optimality of the  simple Bayesian classifier under zero-one loss machine learning",29,pp 103- 130.
[25] Piyush Arora, Akshat Bakliwal and Vasudeva Varma,(2012) "Hindi Subjective Lexicon Generation using WordNet Graph Traversal" In the proceedings of 13th International Conference on Intelligent Text Processing and Computational Linguistics (CICLing ), New Delhi, India
[26] Subhabrata Mukherjee, Pushpak Bhattacharyya,(2012)"Sentiment Analysis in Twitter with Lightweight Discourse Analysis", In Proceedings of the 24th International Conference on Computational Linguistics (COLING 2012.
[27] V. Hatzivassiloglou and K. R. McKeown,(1997), "Predicting the semantic orientation of adjectives," presented at the Proceedings of the eighth conference on European chapter of the Association for Computational Linguistics, Madrid, Spain,.
47